\documentclass[11pt]{JHEP3}

\usepackage{amsmath,amssymb,amsfonts}
\usepackage{epsfig,multicol}

\usepackage{graphicx}% Include figure files
\usepackage{dcolumn}% Align table columns on decimal point
\usepackage{bm}% bold math
\usepackage{epsfig}

\newcommand{\beq}{\begin{equation}}
\newcommand{\eeq}{\end{equation}}
\newcommand{\beqa}{\begin{eqnarray}}
\newcommand{\eeqa}{\end{eqnarray}}
\newcommand{\nn}{\nonumber \\}

\title{Thread-Scalable Evaluation of Multi-Jet Observables}
\author{Walter~T.~Giele\footnote{email: giele@fnal.gov},\ \
  Gerben~C.~Stavenga\footnote{email: stavenga@fnal.gov}\ \
  and\ \ Jan~Winter\footnote{email: jwinter@fnal.gov}\\
  Fermi National Accelerator Laboratory, Batavia, IL 60510, USA}

\preprint{FERMILAB-PUB-10-025-T}

\abstract{
A leading-order, leading-color parton-level event generator is developed
for use on a multi-threaded GPU.
Speed-up factors between 150 and 300
are obtained compared to an unoptimized
CPU-based implementation of the event generator. 
In this first paper we study the feasibility of a GPU-based event
generator with an emphasis on the constraints imposed by the hardware.
Some studies of Monte Carlo convergence and accuracy are presented for 
$PP\rightarrow 2,\ldots,10$ jet observables using 
of the order of $10^{11}$ events.
}
\date{\today}
\begin{document} 

\section{Introduction}
\label{sec:intro}

Leading order (LO) parton-level Monte Carlos (MCs) play a 
prominent role in collider 
phenomenology~\cite{Stelzer:1994ta,Krauss:2001iv,Draggiotis:2002hm,
  Mangano:2002ea,Boos:2004kh,Gleisberg:2008fv}.
As one needs to average the calculation of the observable over many events,
the evaluation time for the event generation is a crucial issue in the development 
of LO parton level MCs.
Furthermore, to make full use of the recent progress in 
the calculation of virtual corrections~\cite{Giele:2008ve,Berger:2008sj,
vanHameren:2009dr}, 
fast tree-level event generators are needed for the calculation 
of the radiative contributions in a next-to-leading order MC.

One can use large-scale grids for the generation of the tree-level events.
Such grids are expensive and need a large infrastructure.
A more preferable solution would be to run the MC on a single, affordable
workstation. As we will show this is possible using a massively parallel GPU. The
NVIDIA Tesla chip is designed for numerical applications~\cite{NVIDIAtesla}
and the CUDA C compiler~\cite{NVIDIAcuda} 
provides a familiar development environment. We will use the Tesla GPU 
throughout the paper.\footnote{We thank the
LQCD collaboration for giving us access to the Tesla GPU processors.}

In this paper we will execute all steps needed for event generation on the GPU.
These steps include the implementation of the unit-weight phase-space
 generator {\sc Rambo}~\cite{Kleiss:1985gy}, 
the evaluation of the strong coupling
and parton density function using LHAPDF~\cite{Whalley:2005nh},
the evaluation of the leading-color $gg\rightarrow 2,\ldots,10$ gluon 
matrix elements at LO and the calculation of some observables.
The CPU is tasked with calculating the distributions using
the event weight and  observables provided by the GPU.
By utilizing memory with a fast access time only, considerable speed-ups
are obtained in the event generation time. This memory is limited
in size, requiring some coding effort.
As the GPU chips are developing fast, we can enhance the 
capabilities of our parton-level generator in accordance. 

In Refs.~\cite{Hagiwara:2009aq,Hagiwara:2009cy} methods have been developed to 
evaluate multi-jet cross sections on GPUs within
the framework of the {\sc Helas} matrix-element evaluator~\cite{Murayama:1992gi},
which forms the basis of the {\sc Madgraph} event generator~\cite{Stelzer:1994ta}. 
The method is based on individual Feynman diagram evaluations. 
As such the scaling with the number
of external particles of the scattering process is faster than factorial. 
Such an algorithm will have limited scalability properties, which cannot
be compensated by deploying a large number of threads.
Instead, an algorithm of polynomial complexity will have excellent
scaling properties; its only limitation is the available fast-access
memory size.
Polynomial algorithms for the evaluation 
of ordered LO multi-parton matrix elements have been formulated
in the form of  Berends--Giele (BG) recursion relations~\cite{Berends:1987me}. 
For a leading-color generator, any Standard Model matrix element 
can be evaluated with an algorithm of polynomial complexity of
degree 4~\cite{Kleiss:1988ne}
or, by using more memory storage, of degree 3~\cite{Draggiotis:1998gr}. 
For any fixed color expansion, the complexity remains polynomial.
Therefore, we will use ordered recursive evaluations 
of the matrix elements instead of Feynman diagram evaluations.

In this paper we present a GPU-based implementation of all basic tools
needed for a LO generator.
In Section~\ref{sec:gen.tes} we discuss the GPU and its
hardware limitations. According to these limitations, we will
determine the optimal running configuration as a function
of the number of gluons. 
The algorithmic implementation of the recursion algorithm and other tools 
such as phase-space generation, 
experimental cuts and parton density functions
are discussed in Section~\ref{sec:spec.tes}.
Finally, in Section~\ref{sec:results} we put all pieces
together and construct the leading-color LO parton-level 
generator capable of generating up to $PP\to10$ jets with sufficient
statistics for serious phenomenology. The conclusions and outlook are
given in Section~\ref{sec:concl}.

\section{Thread-Scalable Algorithms for Event Generators}
\label{sec:gen.tes}

Monte Carlo algorithms belong to a class of algorithms, which can be trivially
parallelized, by dividing the events over the threads.
Optimized for graphics processing, the GPU works by having many threads 
executing essentially the same instructions over different data. 
For a given class of events, e.g.\ $n$-gluon scattering, 
the only difference between the
events is due to the external sources, i.e.\ the momentum and 
polarization four-vectors of each gluon defining
the state of the external gluon.
The recursive algorithm acts on these input sources in an identical manner. 
That is, each thread 
can execute the {\it same}\/ processor instructions 
to calculate the matrix-element weight.

However, because of hardware constraints such a straightforward 
approach is limited by the amount
of available fast access memory. The GPU memory
is independent from the CPU memory and divided into
the off-chip global memory
and the on-chip memory. This distinction is important as the 
off-chip memory is large (of the order of gigabytes)
but slow to access by the threads. Therefore, we want to limit the
access to the global memory by using it only for the transfer of
results to the CPU memory.
The on-chip memory is fast to access, but limited in size (of the order of tens of kilobytes). 
The first on-chip memory structures
are the registers. Each thread has its own registers,
which cannot be accessed by other threads.
These registers are used within the algorithm for variable storage, 
function evaluations, etc.
The other on-chip memory structure is shared memory, which 
is accessible to all the
threads on a multi-processor (MP).
The current GPUs are not yet optimize-able to one event per thread
due to these shared memory and register constraints.
With the next generation of GPUs the shared memory will increase significantly, 
and we will reach the 
point at which we can evaluate one event per thread up to large multiplicities of gluons.

From this discussion the limitations are clear as
each event requires a certain amount of the limited register 
and shared memory.
For the optimal solution, we put
the maximum number of events on one MP, such that
the evaluation does not exceed the available on-chip memory. 
The resulting multiple threads per event
can be used to ``unroll'' do-loops etc., thereby help speed up the
evaluation. This optimal solution is dependent on the rapidly evolving
hardware structure of the GPU chips.

By lowering the number of events per MP below the optimal solution,
the number of available threads per event increases. However, this
will not lead to an effective speed-up of the overall event generation
as the total number of threads per GPU is fixed.
Once the number of events to be used per MP has been 
determined, the GPU evaluation becomes scalable. 
The MC generator now simply scales with the number of available MPs
on the GPU.

\TABLE[!t]{
\label{tab0}
\begin{tabular}{|l|r|r|r|r|r|r|r|r|r|}
\hline&&&&&&&&&\\[-3mm]
$n$           &   4 &  5 &  6 &  7 &  8 &  9 & 10 & 11 & 12 \\[2mm]
\hline&&&&&&&&&\\[-3mm]
events/MP     & 102 & 68 & 48 & 36 & 28 & 22 & 18 & 15 & 13 \\
threads/event &  10 & 15 & 21 & 28 & 36 & 45 & 55 & 66 & 78 \\[2mm]
\hline
\end{tabular}
\caption{The number of $n$-gluon events, which can be simultaneously
  executed on one MP (and is equal to $2048/[n\times(n+1)]$) and the
  number of available threads per event (equal to $n\times(n+1)/2$).
  The total number of events evaluated in parallel on the Tesla chip
  is 30$\times$(events/MP).}}

We will use the current NVIDIA Tesla chip for the numerical evaluations. 
This chip consists of 30 MPs each capable of running up to 1024
threads. Each MP has 16,384 32-bit registers
and an internal shared memory of 16,384 bytes.
Each thread is assigned its own registers from the pool.
Compiling the current MC implementation indicates that 35 registers
per thread are needed.
This gives us an upper
maximum based solely on the use of registers of $16,384/35=468$
threads per MP (each of which could potentially be used to evaluate
one event). The momenta and current storage is of more concern.
As we will see in the next section, 
for the evaluation of the $n$-gluon matrix element, we need to store
$n\times (n+1)/2$ four-vectors in single (float) precision. This requires
$8\times n\times (n+1)$ bytes of shared memory per event on the
MP.\footnote{A bit of calculus shows that when we need to store
  $n\times (n+1)/2$ real-valued four-vectors in single precision we
  require $4\times 4\times n\times (n+1)/2$ bytes of shared memory.}
The resulting maximum number of events per MP as a function 
of the number of gluons is given in Table~\ref{tab0}. Note that up to
44-gluon scattering can be evaluated on the MP (albeit with only one
event per MP). Beyond 44 gluons the shared memory is too small to
store all the required four-vectors.

\section{The Implementation of the Thread-Scalable Algorithm}
\label{sec:spec.tes}

Now that we have determined the optimal running configuration, i.e.\
the number of events per MP, we can implement the algorithm. We will
describe the implementation of the {\sc Threaded EventS Simulator} MC,
which we name {\sc Tess} MC, for the NVIDIA Tesla chip.\footnote{The
  {\sc Tess} MC code can be downloaded from the website: {\tt
    http://vircol.fnal.gov/TESS.html}.} As we have
many threads available per event, we will use these threads to speed
up the MC. In Figure~\ref{fig0} we show the thread usage during
different stages of
the event generator for a $2\rightarrow n$\/ gluon process.
The fraction of the evaluation time spent in each stage depends on the
gluon multiplicity. For 4 gluons, we get 20\%, 20\%, 50\% and
0\% for the {\sc Rambo}, PS-weight, ME-weight and
the epilogue phases, respectively. For 12 gluons, the time consumption
divides up into 2\%, 18\%, 75\% and 4\% for the four phases.

\FIGURE[!t]{
\label{fig0}
\includegraphics[width=1.0\columnwidth]{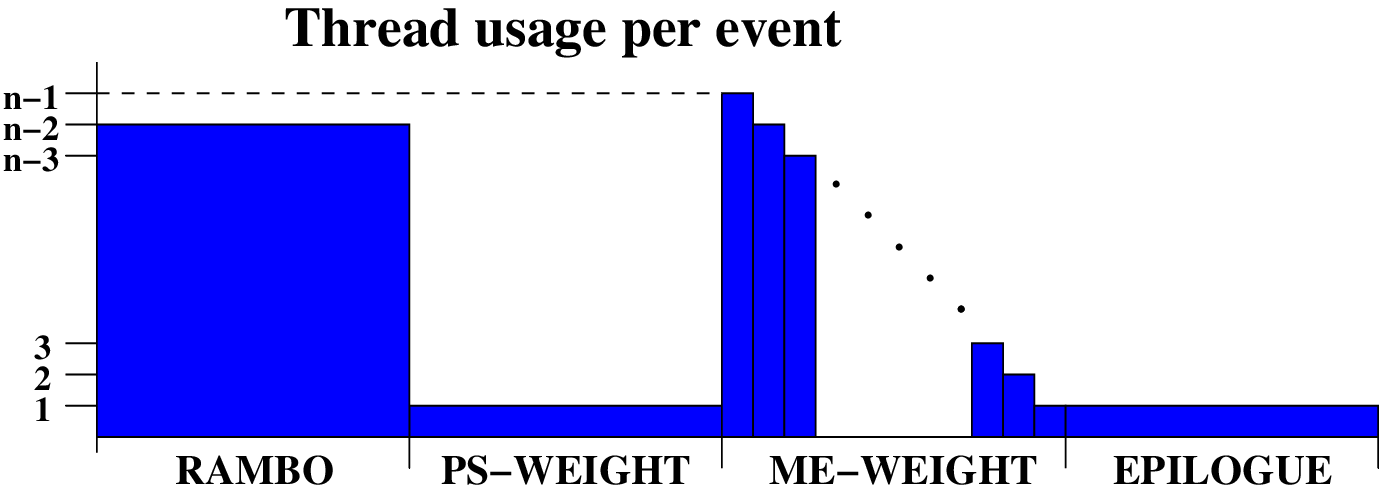}
\caption{
The thread usage for an $n$-gluon event as the algorithm
progresses through the stages of the event generation:
flat phase-space generation, phase-space weight
evaluation (including parton density functions and $\alpha_S$),
matrix-element evaluation and finalization phase.}}

The initialization phase (not shown in Figure~\ref{fig0}) consists of
starting up the kernel on the GPU. This is taken care of by the CUDA
runtime code and does essentially not depend on the
number of threads it has to spawn. However it is a significant part
when the total kernel time is small, as for the 4-gluon case.

The kernel starts initiating the unit-weight phase-space
generator {\sc Rambo}. On the CPU this algorithm grows linearly 
with $n$\/ as we have to construct the $n-2$ outgoing momenta.
On the GPU we can employ $n-2$ threads to simultaneously generate the
outgoing momenta, making the {\sc Rambo} code in practice
independent of $n$.\footnote{The {\sc Rambo} algorithm has some
  summation operations, which grow linearly with $n$, but this time
  scaling is very small compared to the overall evaluation time of the
  {\sc Rambo} algorithm.} 

After the momenta are generated, we have to calculate the strong coupling
constant, the parton density functions and the observables. We also determine,
if the event passes the canonical cuts. If the event fails the cuts, it is
only flagged as such; the matrix-element weight will still be evaluated
as this has no effect on the overall evaluation time. This means one
can deviate from the chosen canonical cuts on the CPU during the histogramming
phase if so desired.
Note that we could in principle generate more events, which pass the cut
before starting the calculation of the matrix-element weights. This
should increase the performance of the Monte Carlo, at the cost of
additional bookkeeping.

The evaluation of the strong coupling constant and parton density functions 
requires special attention. As we have used all shared memory for the
four-vector storage of the gluon currents and momenta, we have to use
the off-chip global memory
to store the parton density and strong coupling constant information
in the form of grids. Furthermore,
interpolation is required between the grid points. To facilitate this, we use 
a special type of memory, the so-called texture memory. This off-chip memory 
was designed for graphics applications and performs hardware interpolations
of the grid.
Specifically, we set up a 1-dimensional grid for the strong
coupling constant. The value of the strong coupling
constant is stored as a function of the renormalization scale at integer values
of the grid. For the 2-dimensional grid used by the parton density
functions, the two dimensions are given by the factorization scale and
the parton fractions.
This parton grid is directly obtained from LHAPDF~\cite{Whalley:2005nh}.
After the grid initialization, the texture memory can be accessed by
the GPU and its hardware will perform the appropriate linear 
interpolation between the grid points when accessing the grid using
non-integer values. This way we have a very fast evaluation of the
strong coupling and parton density functions taking only about $6$\%
and $0.6$\% of the total GPU time for 4-gluon and 12-gluon processes,
respectively.

The four-momenta are generated and the phase-space weight is
determined, hence we have to evaluate the matrix-element weight
next. This happens at the core of the event generator where we use
recursion relations to compute these weights. For this
proof-of-concept program, we decided to use the recursion relation of
Ref.~\cite{Berends:1987me} and restricted ourselves to the case of
pure gluonic cross sections; quarks can be easily added at a later
stage without changing the event generator in a fundamental way.
The recursion relations we employ are given by
\begin{multline}\label{BG-recursion}
J_{\mu}[m,\ldots,n]\;=\;\frac{1}{K[m,\ldots,n]^2}\;
\Bigg(\,\sum_{i=m}^{n-1} \Big[J[m,\ldots,i], J[i+1,\ldots,n]\Big]_{\mu}\\
+\;\sum_{i=m}^{n-2}\sum_{j=i+1}^{n-1} 
\Big\{J[m,\ldots,i], J[i+1,\ldots,j], J[j+1,\ldots,n]\Big\}_{\mu}\Bigg)\ ,
\end{multline}
where $J_{\mu}[m,\ldots,n]$ is a conserved four-vector current
depending on the external gluons $\{m,\ldots,n\}$. 
Furthermore, we have used the shorthand notations
\begin{equation}
\begin{split}
K^{\mu}[m,\ldots,n]\;&=\;\sum_{i=m}^n k_i^{\mu}\ ,\\[3mm]
\Big[J[\{a\}], J[\{b\}]\Big]_{\mu}\,&=\;
2\, \Big(J[\{a\}]\cdot K[\{b\}]\Big)\, J_{\mu}[\{b\}]
-2\, \Big(K[\{a\}]\cdot J[\{b\}]\Big)\, J_{\mu}[\{a\}]\\
&+\Big(J[\{a\}]\cdot J[\{b\}]\Big)\, \Big(K_{\mu}[\{a\}]-K_{\mu}[\{b\}]\Big)\ ,\\[3mm]
\Big\{J[\{a\}],J[\{b\}],J[\{c\}]\Big\}_{\mu}\,&=\;2\,\Big(J[\{a\}]\cdot J[\{c\}]\Big) J_{\mu}[\{b\}]\\
&-\Big(J[\{a\}]\cdot J[\{b\}]\Big)\, J_{\mu}[\{c\}]
-\Big(J[\{c\}]\cdot J[\{b\}]\Big)\, J_{\mu}[\{a\}]\ ,
\end{split}
\end{equation}
where the external gluon labeled $i$\/ has momentum $k_i^{\mu}$ and
polarization state $J^{\mu}[i]$.
These four-vectors form the initial conditions for the recursion
relation. In addition to the $n$\/ momenta, the recursion relation
requires $n\times(n-1)/2$ four-vector currents to be stored giving a
total storage of $n\times(n+1)/2$ four-currents per event.

The recursion relations have a polynomial complexity of order
$n^4$ for calculating the currents~\cite{Kleiss:1988ne}.
By exploiting the available threads for each event,
we can reduce the algorithmic complexity of the BG recursion relation.
The relation is easily thread-able,
which enables us to lower the polynomial scaling of the evaluation
time of the recursion relation to $n^3$. A full recursion for an
$n$-gluon process is completed in $n-1$ steps. In the first step, we
use $n-1$ threads to calculate the polarization vectors
$\{J[2],\ldots,J[n]\}$ needed as a starting point in the recursion
relation. We choose each polarization vector as a random unit vector
orthogonal to the respective gluon momentum. 
By doing this, instead of employing the conventional helicity vectors, we obtain
real-valued currents. This avoids complex multiplications and
reduces the shared-memory usage, resulting in a significant time gain.

After the 1-currents have been determined, we use $n-2$ threads to
calculate the 2-currents $\{J[2,3],J[3,4],\ldots,J[n-1,n]\}$. We
continue with the $n-1$ steps until we have determined
$J[2,3,\ldots,n]$ at which point we can calculate the ordered
amplitude and, hence, the matrix-element weight. Note that because we
make use of the multiple threads we have reduced the computational
effort from ${\cal O}(n^4)$ to ${\cal O}(n^3)$ complexity.

In principle we may be able to improve even further. The initial
${\cal O}(n^4)$ growth of the one-threaded recursion relation to
calculate the $J[2,3,\ldots,n]$ current can be reduced by rewriting
the recursion relation as
\begin{equation}
J_{\mu}[m,\ldots,n]\;=\;\sum_{i=m}^{n-1}\left[
 \Big(W[i+1,\ldots,n]\cdot J[m,\ldots,i]\Big)_{\mu}
-\Big(W[m,\ldots,i]\cdot J[i+1,\ldots,n]\Big)_{\mu}\right]\ ,
\end{equation}
where the tensor $W_{\mu\nu}$ is defined as
\begin{equation}
\begin{split}
W_{\mu\nu}[m,\ldots,n]\;&=\;
2\,J_{\mu}[m,\ldots,n]\, K_{\nu}[m,\ldots,n]-
K_{\mu}[m,\ldots,n]\, J_{\nu}[m,\ldots,n]\\
&+\sum_{i=m}^{n-1}\Big(J_{\mu}[m,\ldots,i]\, J_{\nu}[i+1,\ldots,n]
-J_{\mu}[i+1,\ldots,n]\, J_{\nu}[m,\ldots,i]\Big)\ .
\end{split}
\end{equation}
By undoing the nested summations in the second term of
Eq.~\ref{BG-recursion} we have lowered the complexity of the algorithm
to ${\cal O}(n^3)$. However, this is only achieved at the cost of
using significantly more storage. For each event, one would have to
store $n\times (n-1)/2$ $4\times4$-tensors in addition to the $n\times
(n+1)/2$ momenta and current four-vectors. Up to $n\approx10$
the extra work of doing matrix multiplications together with the 
fact that the relative pre-factor of the $n^4$-algorithm is small,
$1/4$, compared to the $n^3$-algorithm actually make the
$n^3$-algorithm slower than the $n^4$-algorithm. Moreover, the extra
storage demand does not make the $n^3$-algorithm attractive for our
GPU implementation.

From the current for $n-1$ gluons we then obtain the amplitude for the
$n$-gluon matrix element by putting the off-shell leg on-shell,
contracting in with the final polarization vector and symmetrizing
over the gluons in the current. Specifically,
\begin{equation}\label{amplitude}
\mathcal{A}(1,\ldots,n)\;=\;
\sum_\pi^{(n-1)!} \mbox{Tr}
\Big[T^{a_{\pi_1}}\ldots T^{a_{\pi_{n-1}}}T^{a_n}\Big]\, 
m(\pi_1,\ldots,\pi_{n-1},n)\ ,
\end{equation}
with
\begin{equation}
m(\pi_1,\ldots,\pi_{n-1},n)\;=\;
\left.\Big(J[\pi_1,\ldots,\pi_{n-1}]\cdot J[n]\Big)
\times K^2[1,\ldots,n-1]
\right\rfloor_{K[1,\ldots,n]=0}\ .
\end{equation}
Notice that for a given phase-space point, we have to perform the
permutation sum requiring $(n-1)!$ steps to arrive at the full
amplitude. This would immediately lead to a factorial growth
in the computer time.  We can circumvent the super-exponential sum
over permutations in  Eq.~\ref{amplitude}. In the leading-color
approximation this is easily accomplished and the color-summed,
squared amplitude is given by
\begin{equation}
\left|\mathcal{A}(1,\ldots,n)\right|^2\;\sim\;N_\mathrm{C}^{n-2}
\left(N_\mathrm{C}^2-1\right)\left(
\sum_\pi^{(n-1)!}\left|m(\pi_1,\ldots,\pi_{n-1},n)\right|^2 
+ {\cal O}\left(\frac{1}{N_\mathrm{C}^2}\right)\right)\ .
\end{equation}
As we will use this matrix element in a $2\rightarrow n-2$
gluon-scattering phase-space integration, we can use the symmetry of
the final state to remove the permutation sum over the ordered
amplitudes. In detail,
\beqa\lefteqn{
d\,\sigma(PP\rightarrow n-2\ \mbox{jets})} \nn
&=&
\int d\,x_1\ d\,x_2\ \frac{F_g(x_1)F_g(x_2)}{4\,p_1\cdot p_2}\frac{1}{(n-2)!}
\int d\,\Phi(p_1p_2\rightarrow p_3\cdots p_n)
\sum_\pi^{(n-1)!}\left|m(\pi_1,\ldots,\pi_{n-1},n)\right|^2 \nn
&=&
\int d\,x_1\ d\,x_2\ \frac{F_g(x_1)F_g(x_2)}{4\,p_1\cdot p_2}\, (n-1)
\int d\,\Phi(p_1p_2\rightarrow p_3\cdots p_n)
\left|m(1,\sigma_2,\ldots,\sigma_n)\right|^2\ ,
\eeqa
where $p_1=x_1\,P_1$, $p_2=x_2\,P_2$, the parton density function 
is given by $F_g(x)$, $d\,\Phi$ is
the phase-space integration measure and $\{\sigma_2,\ldots,\sigma_n\}$
is a permutation of the list $\{2,\ldots,n\}$ assigned randomly for
each MC phase-space point evaluation.

Eventually, in the very last step of our threaded event simulation all
the results are put together and returned to the CPU for processing.

By using the {\sc Tess} MC, we can
evaluate the differential $n$-jet cross sections in the leading-color
approximation.
The algorithm is of polynomial complexity and scales
as $n^3$ with the number $n$\/ of gluons.

\section{A Numerical Study of the Threaded Events Simulator}
\label{sec:results}

\FIGURE[!t]{
\label{fig1}
\centerline{
  \includegraphics[clip,width=0.48\columnwidth]{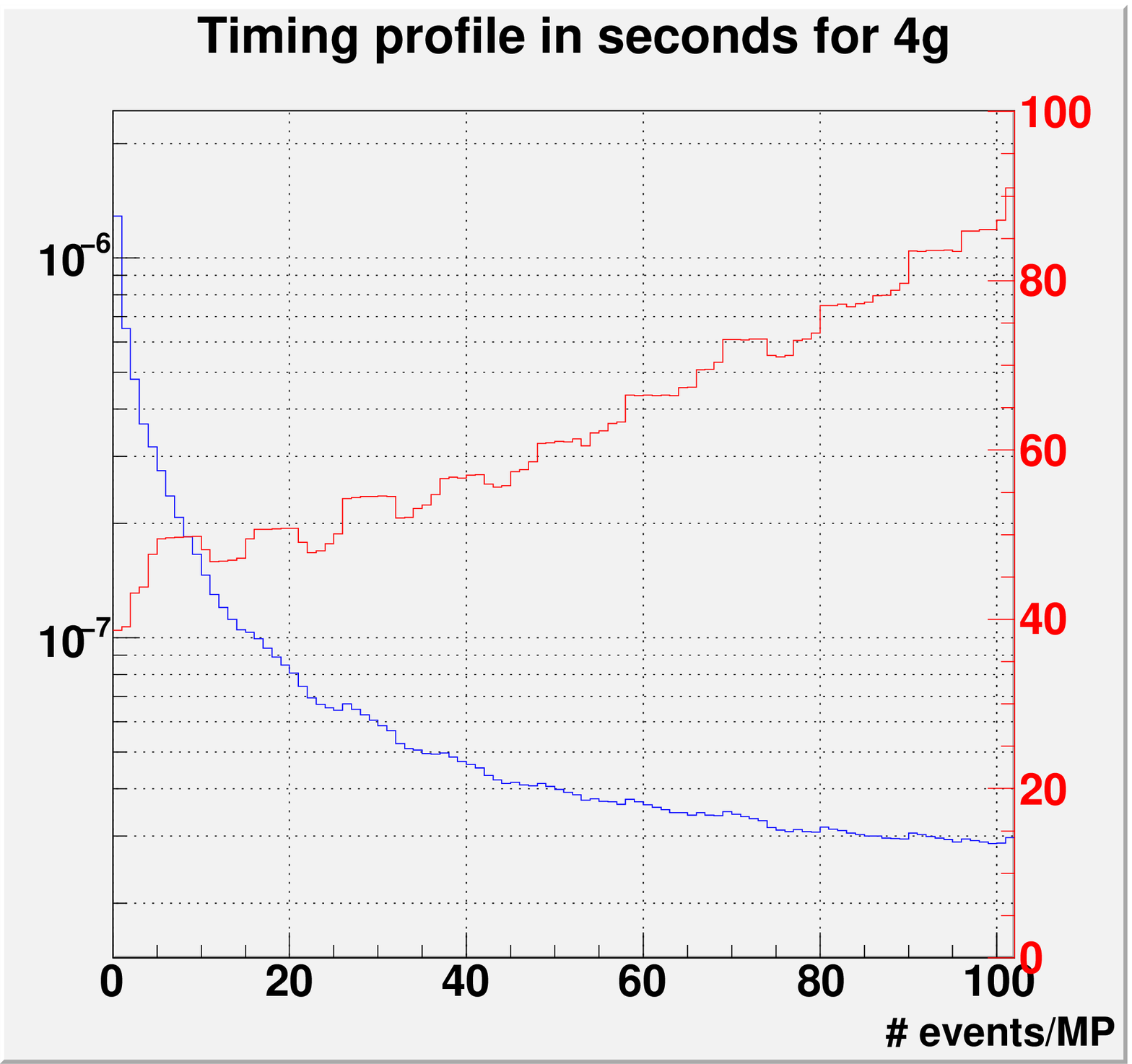}
  \includegraphics[clip,width=0.48\columnwidth]{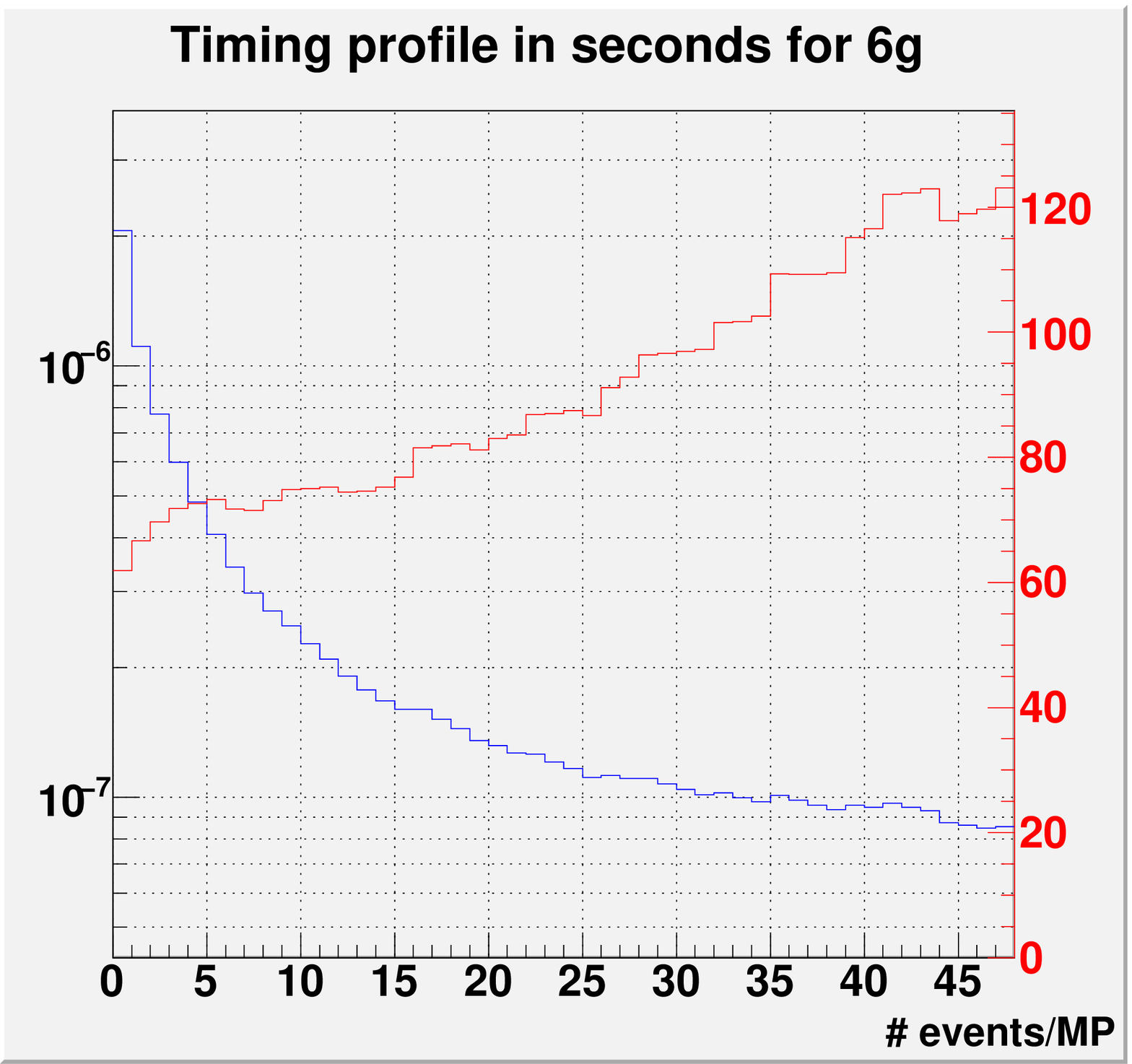}}\\[1mm]
\centerline{
  \includegraphics[clip,width=0.48\columnwidth]{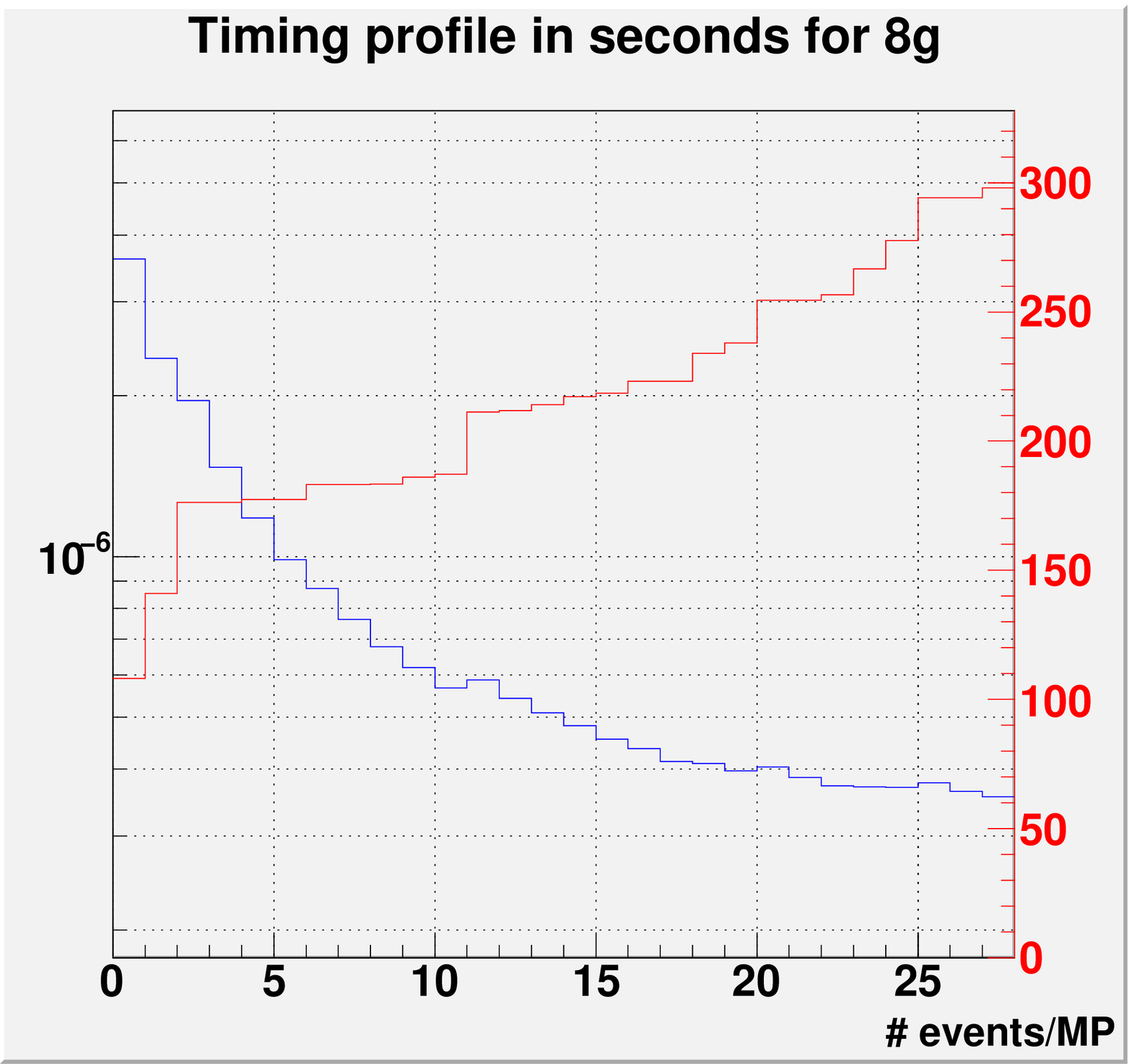}
  \includegraphics[clip,width=0.48\columnwidth]{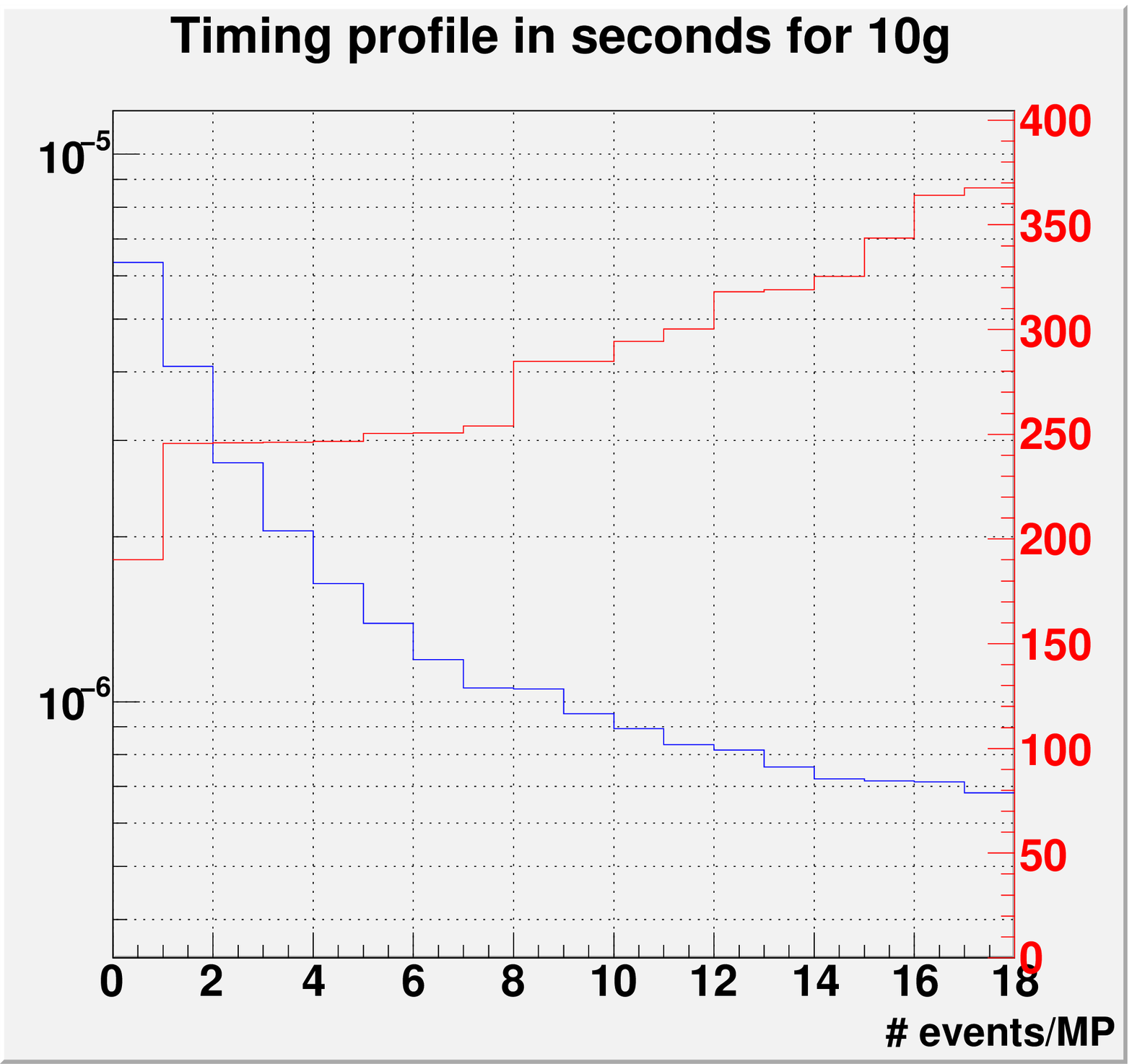}}
\caption{The horizontal axis is the number of events per MP in a
  sweep, giving a total number of 30$\times$(events per MP) evaluated
  events per sweep. The red curves used together with the vertical
  axes on the right indicate the total GPU time in seconds for
  1,000,000 sweeps. The blue curves depict the evaluation time of one
  event in seconds as labeled by the vertical axes on the left.}}

The first issue to study is the timing behavior of the {\sc Tess} MC. We show
our results in Figure~\ref{fig1} where we plot the GPU timing as a
function of events per MP for several gluon multiplicities.
Each MP will evaluate in a sweep a number of events in
parallel using the threads. In principle the sweep time should be
independent of the number of events evaluated by each MP as long as
the shared-memory constraints are not exceeded, cf.\ Table~\ref{tab0}.
However, we have to execute a substantial amount of transcendental
function calls per event, which induces some queuing at the special
function units each MP uses for evaluating these functions. This
queuing effect will increase as the number of events per MP rises and,
hence, lead to a slower execution of the sweep. In Figure~\ref{fig1}, one
can see this complicated timing behavior, which is controlled by the GPU
hardware. As discussed the overall evaluation time increases with the
number of events per MP, see the red curves in the plots. In fact, the
increase of the overall evaluation time is overcome by the gain we
achieve in evaluating more events per MP. The more relevant quantity
therefore is the evaluation time per event, defined as the GPU
evaluation time divided by the total number of generated events. As
clearly indicated by the blue curves in Figure~\ref{fig1}, the time
consumption per event steadily decreases as the number of events per
MP increases. The best performance will be achieved by using the
maximal number of events available per MP.

\TABLE[!t]{
\label{tab1}
\begin{tabular}{|c|c|c||c|c||c|}
\hline&&&&&\\[-3mm]
$n$ & $T_n^\mathrm{GPU}$ (seconds) & $P_n(3)$ &
      $T_n^\mathrm{CPU}$ (seconds) & $P_n(4)$ &
      $G_n$\\[2mm]
\hline&&&&&\\[-3mm]
4  & $2.975\times 10^{-8}$ &      & $8.753\times 10^{-6}$ &      & 294\\
5  & $4.438\times 10^{-8}$ & 0.91 & $1.247\times 10^{-5}$ & 0.87 & 281\\
6  & $8.551\times 10^{-8}$ & 1.03 & $1.966\times 10^{-5}$ & 0.93 & 230\\
7  & $2.304\times 10^{-7}$ & 1.19 & $3.047\times 10^{-5}$ & 0.96 & 132\\
8  & $3.546\times 10^{-7}$ & 1.01 & $4.736\times 10^{-5}$ & 0.98 & 133\\
9  & $4.274\times 10^{-7}$ & 0.94 & $7.263\times 10^{-5}$ & 0.99 & 170\\
10 & $6.817\times 10^{-7}$ & 1.05 & $1.044\times 10^{-4}$ & 0.99 & 153\\
11 & $9.750\times 10^{-7}$ & 1.02 & $1.529\times 10^{-4}$ & 1.00 & 157\\
12 & $1.356\times 10^{-6}$ & 1.02 & $2.129\times 10^{-4}$ & 1.00 & 158\\[2mm]
\hline
\end{tabular}
\caption{
The GPU and CPU evaluation times per event, $T_n^\mathrm{GPU}$
and $T_n^\mathrm{CPU}$, given as a function of the number $n$\/
of gluons for $gg\to(n-2)\,g$ processes. The polynomial scaling
measures are also shown, for the GPU, $P_n(3)$, and for the CPU,
$P_n(4)$. The $P_n(m)$ are defined as
$P_n(m)=[(n-1)/n]\times\sqrt[m]{T_n/T_{n-1}\;}$. The rightmost
column finally displays the gain $G=T_n^\mathrm{CPU}/T_n^\mathrm{GPU}$.}}

Now that we have determined the optimal running conditions, we give in Table~\ref{tab1}
the evaluation time per event on the GPU compared to the evaluation
time of the same algorithm when executed
on the CPU. As can be seen the speed-up in evaluation time is substantial, 
ranging from almost
a factor of 300 for 4-gluon processes to a factor of around 150 for
12-gluon processes.
Note that the speed-up is completely due to the fact that we evaluate
in parallel 3060 and 390 events for the 4- and 12-gluon case,
respectively. We have also tested that running the events sequentially
on the GPU using only one event per sweep results in an event evaluation time,
which is slower than the CPU evaluation time. In particular, we found
factors of 10 and 2 for the 4- and 12-gluon computations, respectively.
Because of the substantial time gains, a single GPU
can replace a large grid of hundreds of CPUs.

Also of interest is the scaling behavior of the algorithm.
As expected, on the CPU it is simply polynomial scaling with a factor of 4 
in the limit of a large number of gluons.
We see from the table that this scaling is setting in quickly. 
The GPU algorithm scales with a factor 
of 3 as discussed in Section~\ref{sec:spec.tes}. However, as the number of gluons increases, the number of events
per MP decreases. This makes the timing more dependent on specific hardware issues. As
can be seen from Table~\ref{tab1} the polynomial 
scaling is trending towards a factor of 3. 

\FIGURE[!]{
\label{fig2}
\centerline{
  \includegraphics[clip,width=0.46\columnwidth]{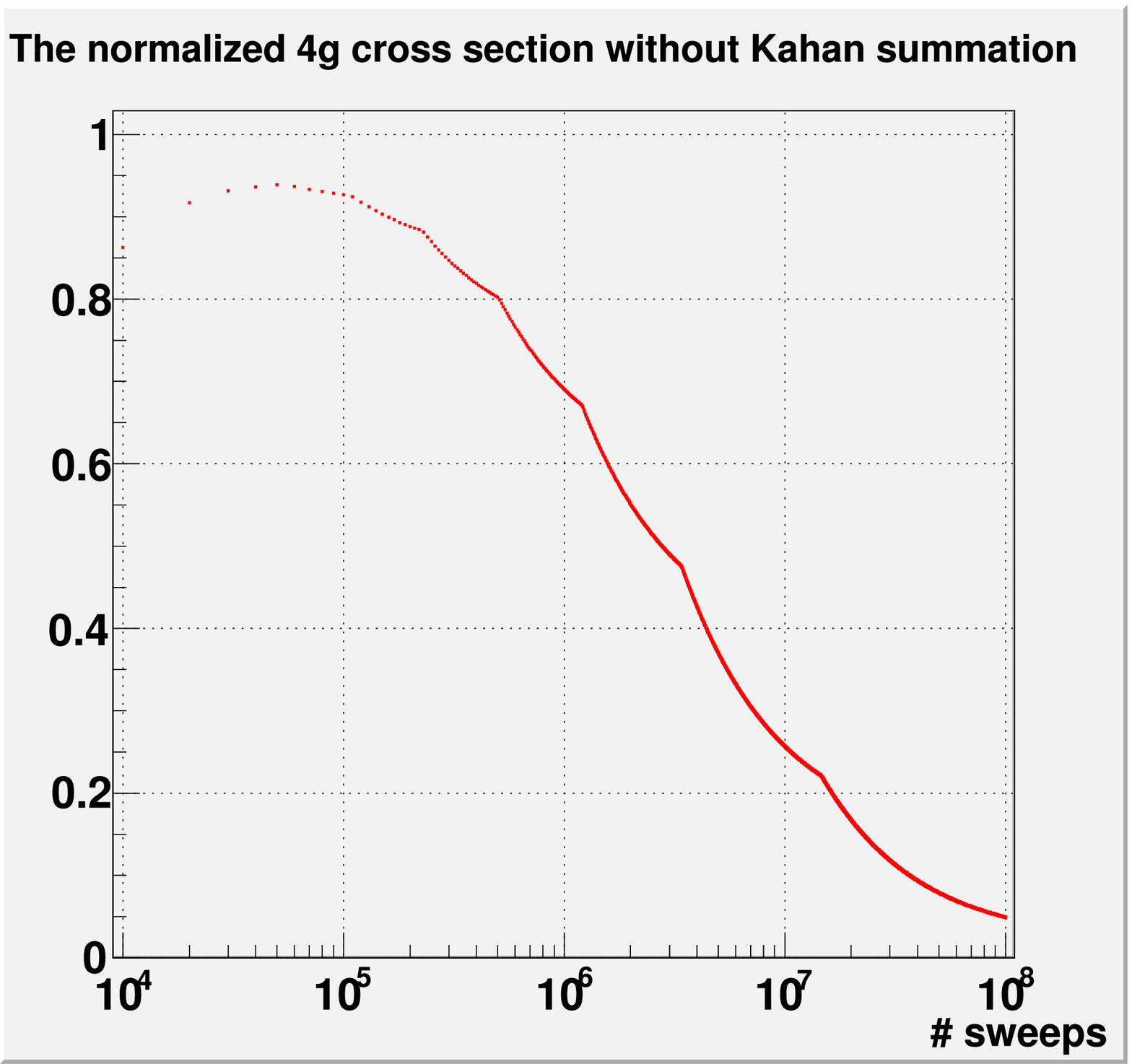}
  \includegraphics[clip,width=0.46\columnwidth]{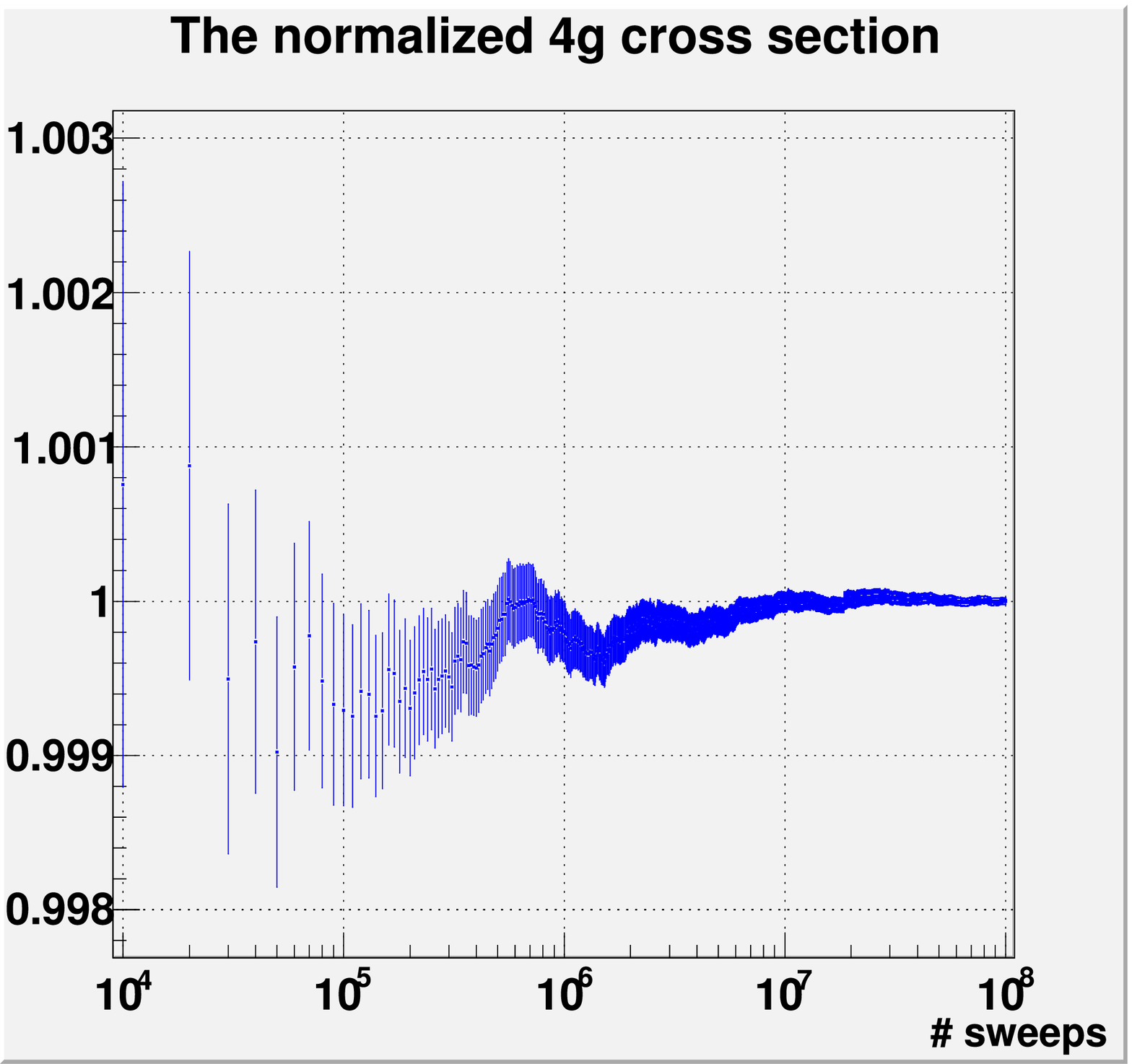}}\\[1mm]
\centerline{
  \includegraphics[clip,width=0.46\columnwidth]{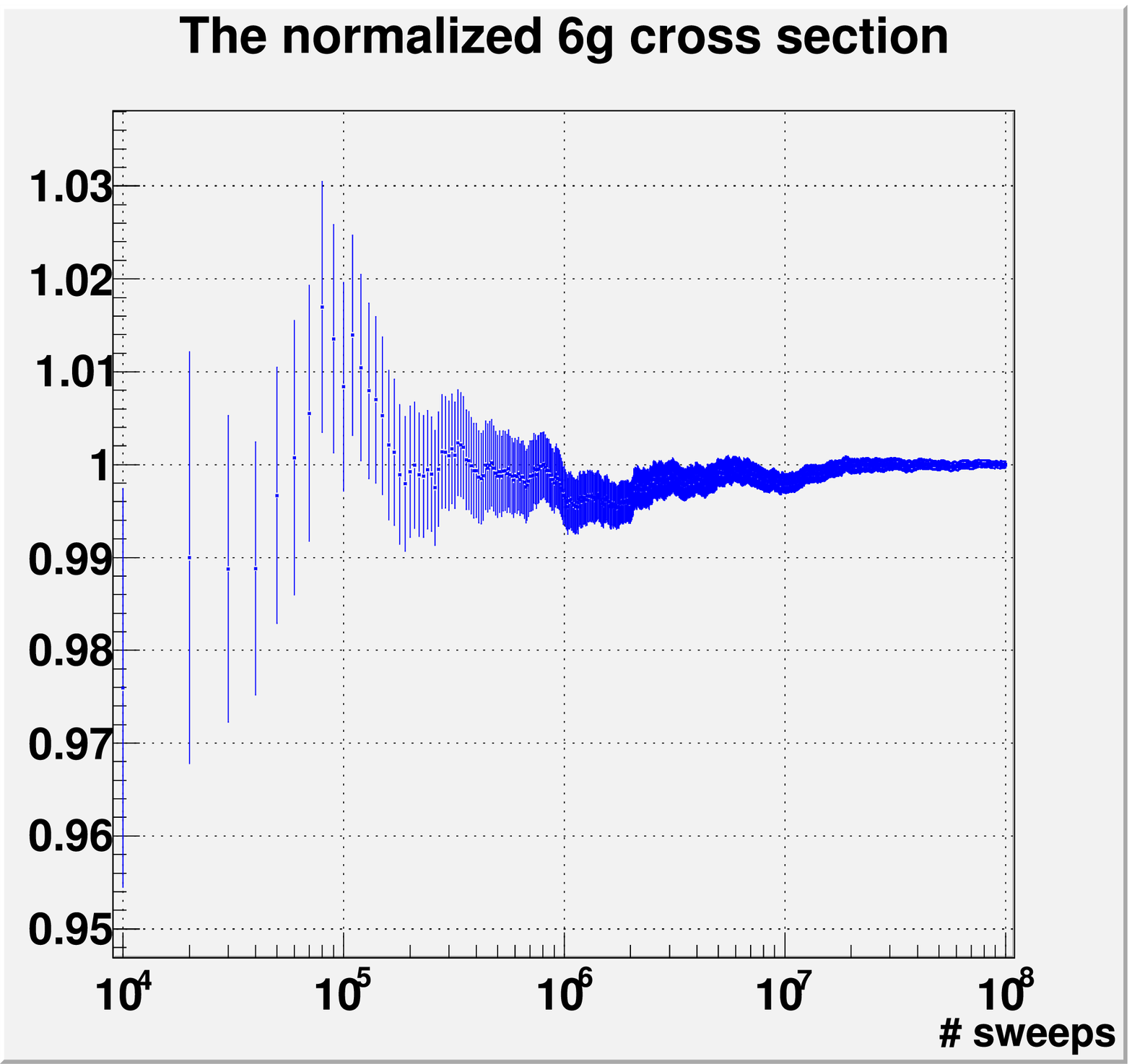}
  \includegraphics[clip,width=0.46\columnwidth]{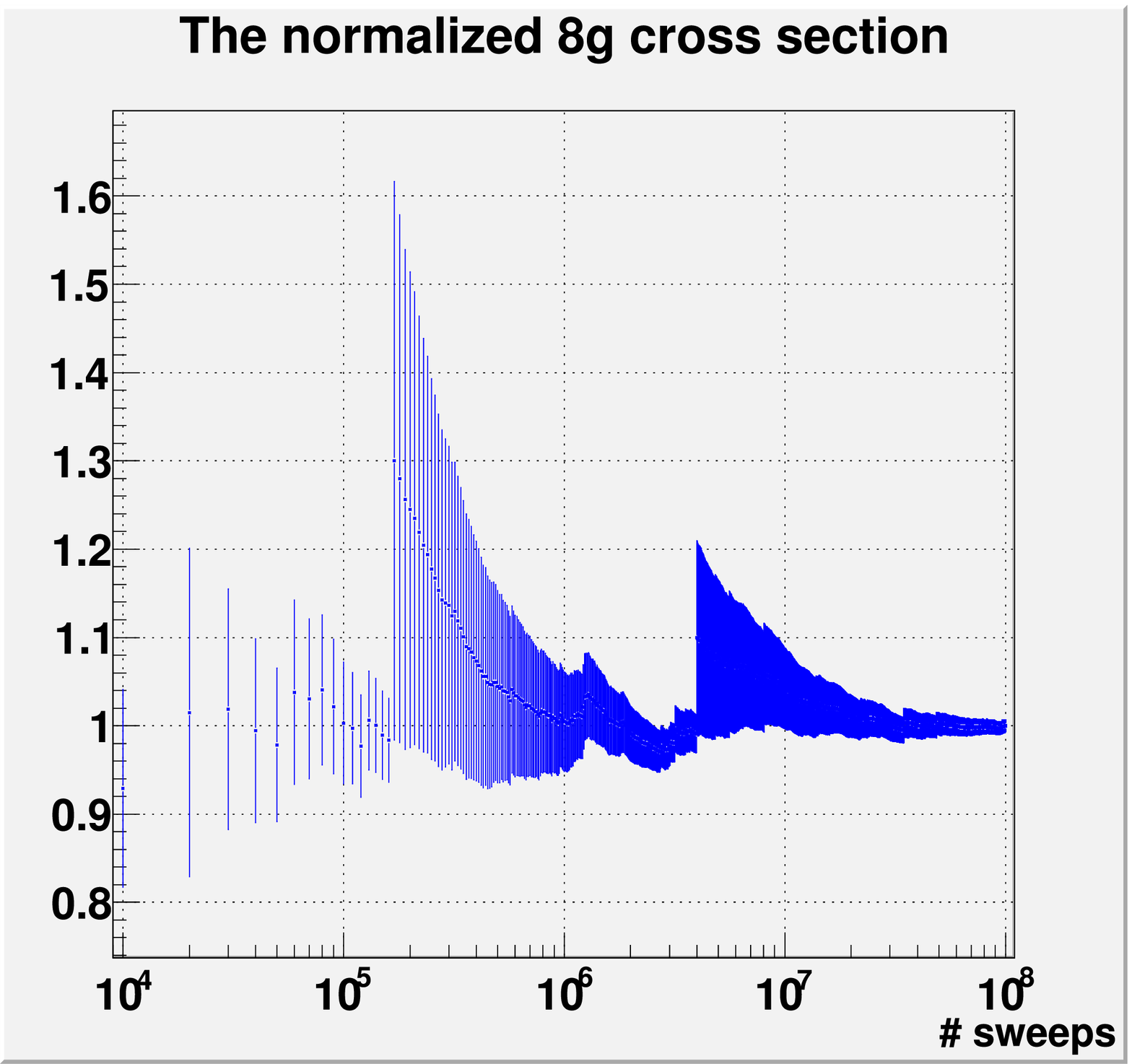}}\\[1mm]
\centerline{
  \includegraphics[clip,width=0.46\columnwidth]{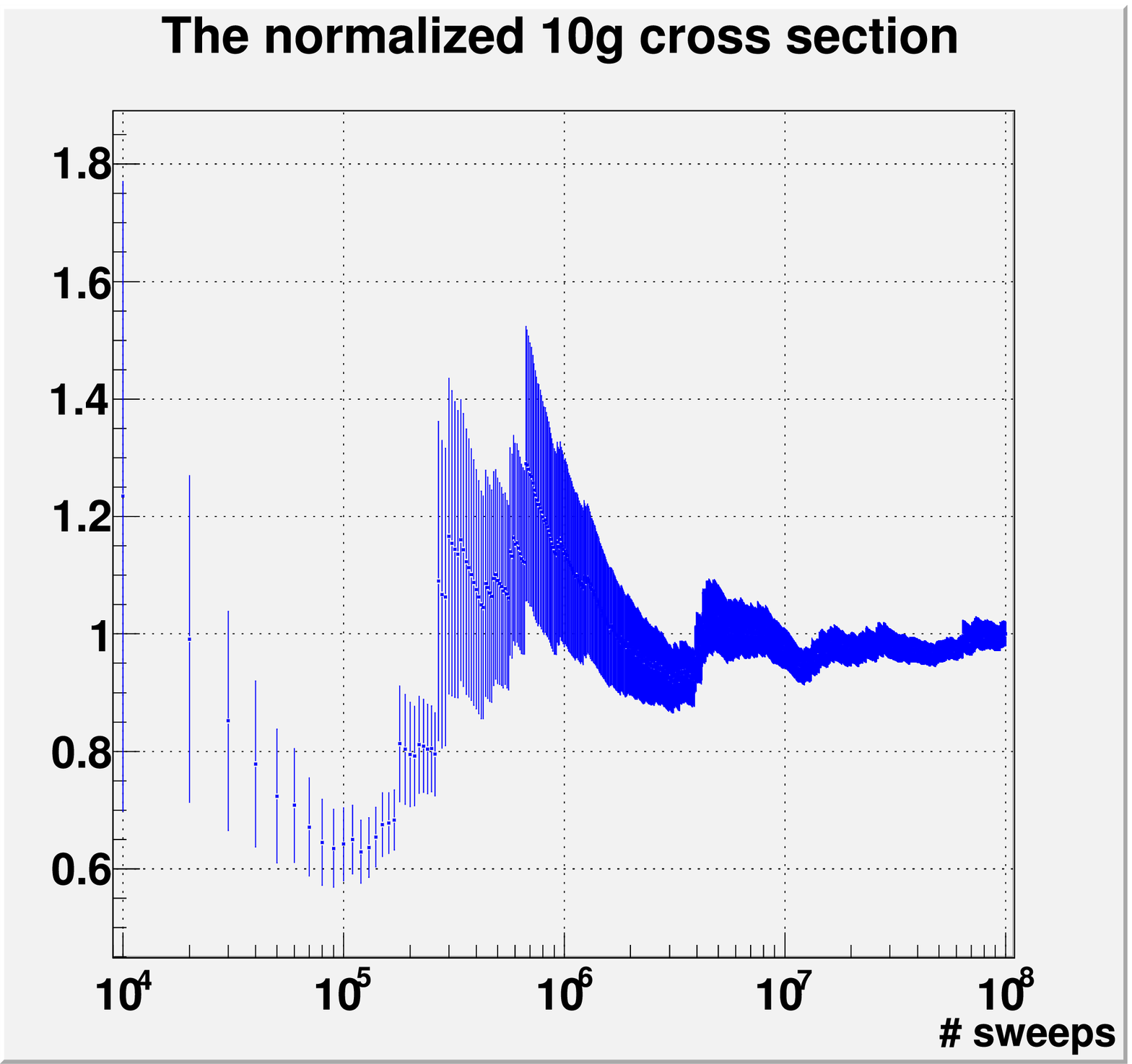}
  \includegraphics[clip,width=0.46\columnwidth]{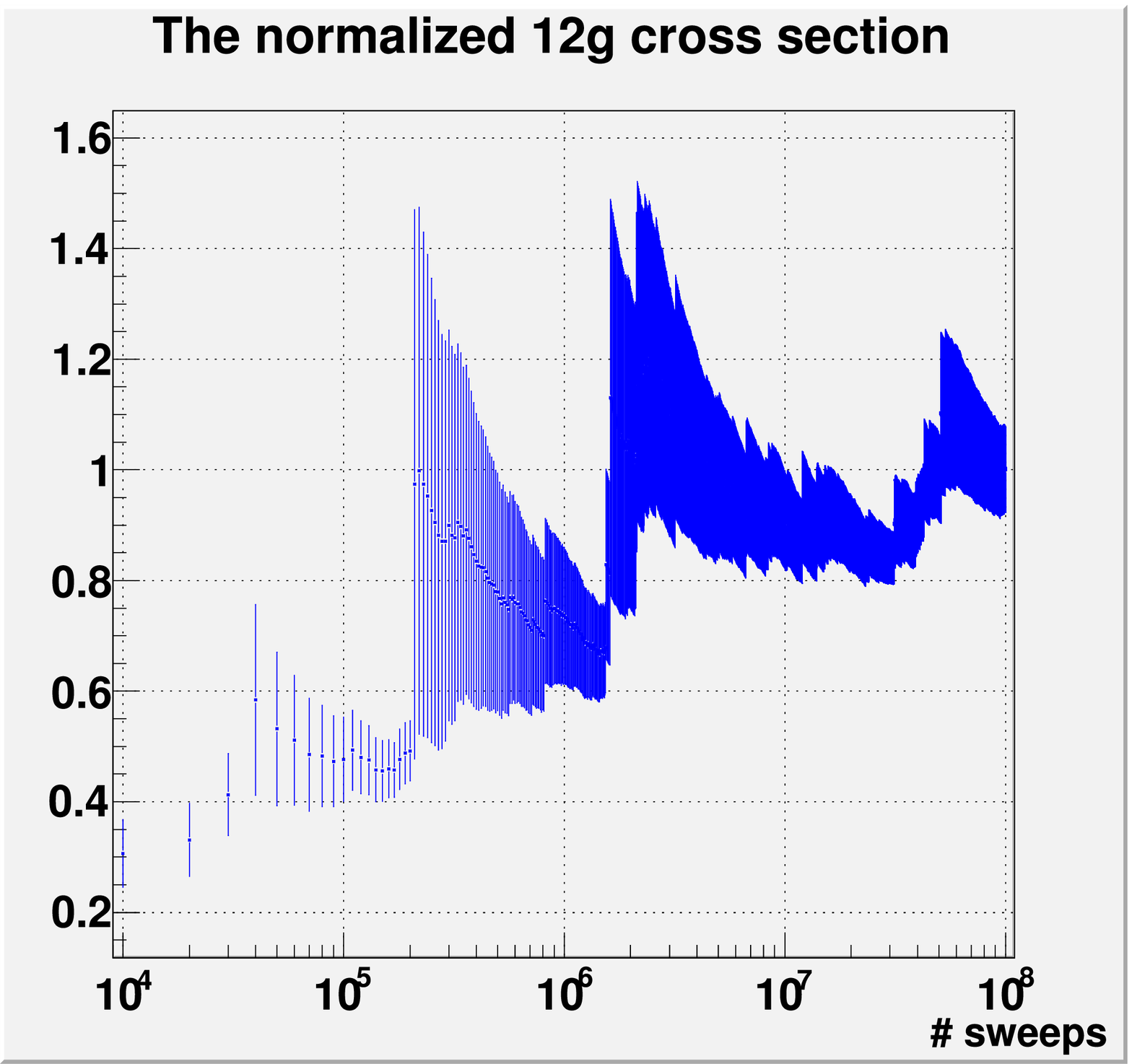}}
\caption{
The number of sweeps versus several $gg\to(n-2)\,g$ cross sections
normalized to their respective best cross section estimates as
given in Table~\ref{tab2}. The error is the mean standard deviation.}}

Given the fast evaluation of events, we can easily generate ${\cal O}(10^{11})$
events for the calculation of the LO cross sections. With these large
numbers of generated events, one has to carefully consider the
performance of the random number generators. In our case this should
cause no issues, since the number of generated random numbers is of
the order of the square root of the generator's cycle length. However,
as we average over ${\cal O}(10^{11})$ numbers, care has to be taken
concerning the loss of precision, which would result in a systematic
underestimation of the cross section. This is demonstrated in the
first graph of Figure~\ref{fig2} where we have used a single-precision
summation to calculate the 4-gluon cross section. As can be seen the
effect becomes dramatic as the number of sweeps is rising and we end
up with a totally wrong determination of the cross section. We
avoid this problem by using the Kahan summation
algorithm~\cite{Kahan}. All other graphs of the figure are produced by
following this procedure. These additional graphs display the
convergence of the cross section estimate including its mean standard
deviation as a function of the number of GPU sweeps. The vertical axis
has been normalized to the respective best estimate of the cross
section; all of which are listed in Table~\ref{tab2}. For this
study, we have used {\sc Rambo} as the momenta generator, therefore, a
severe under-sampling of small phase-space regions with large weights
may occur especially for larger gluon multiplicities. Because the
{\sc Rambo} phase-space generation is flat and does not reflect the
scattering amplitudes' strong dipole structure, such under-sampling
effects are expected and cause the peaking behavior of our cross
section estimates. Even with ${\cal O}(10^{10})$ phase-space points an
estimate of the 12-gluon cross section using the {\sc Rambo} event generator
is quite unreliable and the mean standard deviation error estimate
does not fully reflect the true uncertainty. In a further development
step, one may implement a phase-space generator like
{\sc Sarge}~\cite{Draggiotis:2000gm}, which is capable of adapting to the
QCD antenna structures as occurring in the matrix elements. As pointed
out in Ref.~\cite{Gleisberg:2008fv}, this would resolve the
phase-space integration issues we have seen here.

\TABLE[!t]{
\label{tab2}
\begin{tabular}{|c|r|r|l||r|} 
\hline&&&&\\[-3mm]
$n$ & \multicolumn{1}{c|}{$\sigma_n$ (pb)} &
\multicolumn{1}{c|}{$N_\mathrm{generated}$} &
\multicolumn{1}{c||}{$N_\mathrm{accepted}$} &
\multicolumn{1}{c|}{$\sigma_n^\textsc{Comix}$ (pb)}\\[2mm]
\hline&&&&\\[-3mm]
4  & $(2.32421\pm 0.00047)\times 10^8$ & $3.06\times 10^{11}$ & $1.96848\times 10^{11}$ & $(2.3283\pm 0.0023)\times 10^8$ \\
5  & $(1.4353\pm 0.0011)\times 10^7$   & $2.04\times 10^{11}$ & $1.12939\times 10^{11}$ & $(1.4355\pm 0.0014)\times 10^7$ \\ 
6  & $(2.84780\pm 0.00096)\times 10^6$ & $1.44\times 10^{11}$ & $6.98918\times 10^{10}$ & $(2.8560\pm 0.0030)\times 10^6$ \\
7  & $(6.356\pm  0.012)\times 10^5$    & $1.08\times 10^{11}$ & $4.49985\times 10^{10}$ & $(6.408\pm 0.015)\times 10^5$   \\
8  & $(1.608\pm 0.011)\times 10^5$     & $8.40\times 10^{10}$ & $2.93316\times 10^{10}$ &                                 \\
9  & $(4.38\pm  0.11)\times 10^4$      & $6.60\times 10^{10}$ & $1.88182\times 10^{10}$ &                                 \\
10 & $(1.193\pm  0.024)\times 10^4$    & $5.40\times 10^{10}$ & $1.22356\times 10^{10}$ &                                 \\
11 & $(3.550\pm  0.020)\times 10^4$    & $4.50\times 10^{10}$ & $7.88017\times 10^9   $ &                                 \\
12 & $(9.64361\pm  0.74)\times 10^3$   & $3.90\times 10^{10}$ & $5.13041\times 10^9   $ &                                 \\[2mm]
\hline
\end{tabular}
\caption{
The cross sections $\sigma_n$ for $gg\to(n-2)\,g$ and their mean
standard deviations in pb as calculated by the {\sc Tess} MC using
$10^9$ sweeps. The two center columns show the total number of
generated events and the number of events passing the jet cuts.
For comparison, the cross sections $\sigma_n^\textsc{Comix}$
in pb as computed by {\sc Comix} considering the full-color dependence
are also given.}}

\FIGURE[!]{
\label{fig3}
\centerline{
  \includegraphics[clip,width=0.890\columnwidth]{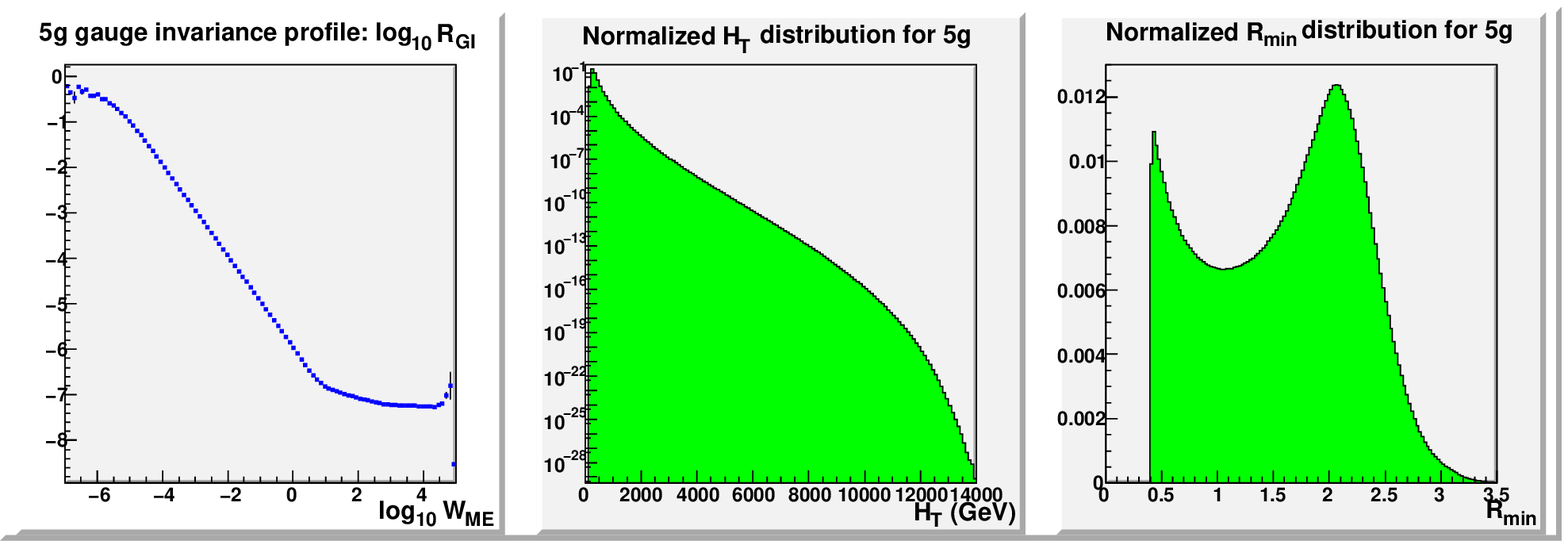}
}\\[1mm]
\centerline{
  \includegraphics[clip,width=0.890\columnwidth]{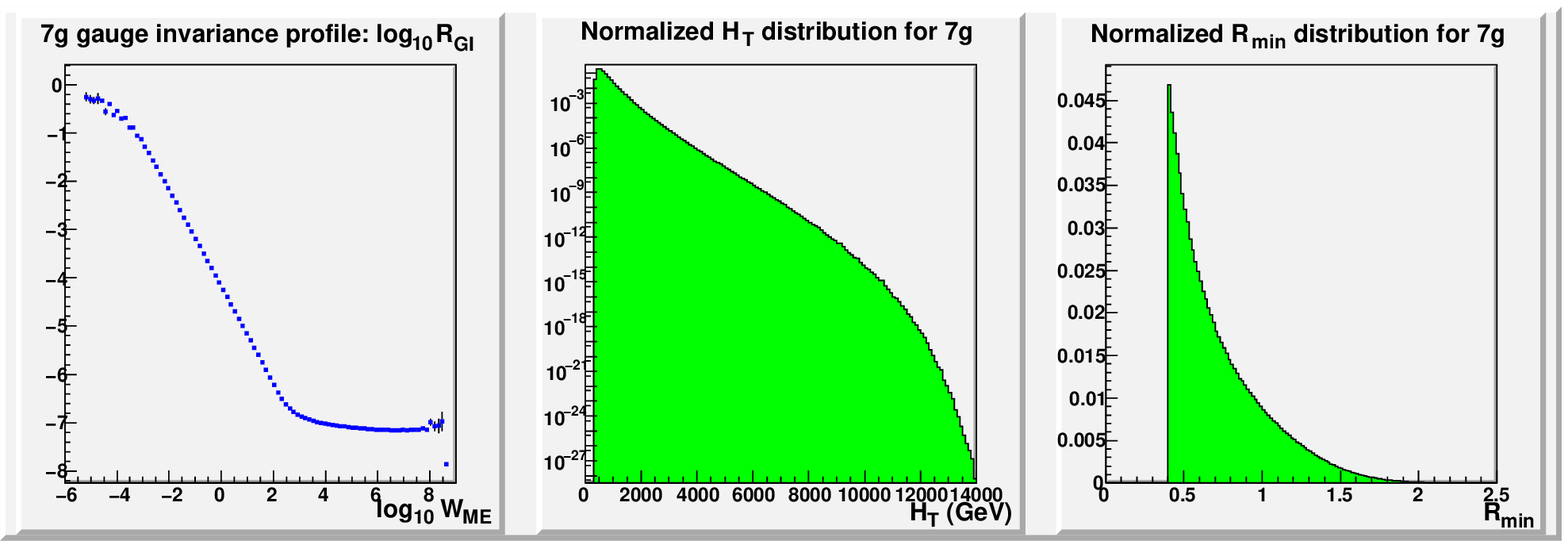}
}\\[1mm]
\centerline{
  \includegraphics[clip,width=0.890\columnwidth]{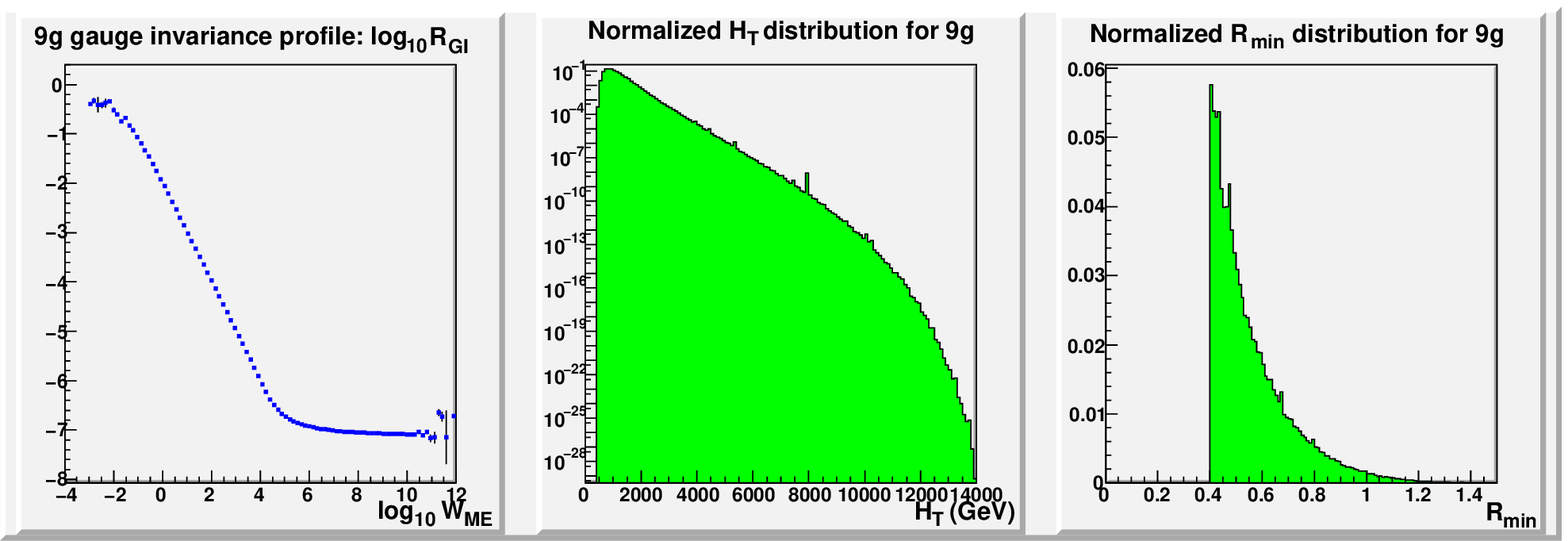}
}\\[1mm]
\centerline{
  \includegraphics[clip,width=0.890\columnwidth]{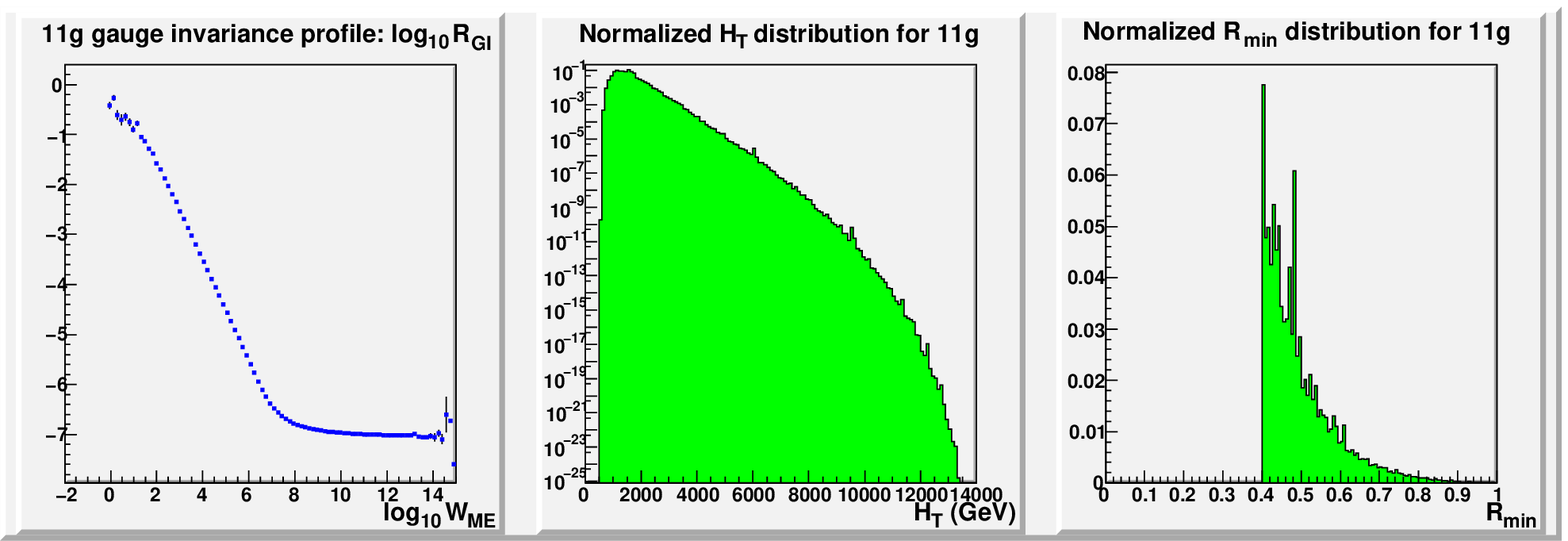}
}\\[1mm]
\caption{
Left panels: the profile plots of the relative gauge invariance as a
function of the decimal logarithm of the matrix-element weight,
$\log_{10}W_\mathrm{ME}$; center
panels: the normalized $H_T$ distributions and right panels:
the normalized minimum $R$-separations between pairs of jets. All of
which is shown for $gg\to(n-2)\,g$ scatterings at a 14 TeV
center-of-mass energy for $n=5,7,9,11$.}}

The convergence issues reflected in Figure~\ref{fig2} should be taken
into account when interpreting the uncertainties of our best cross
section estimates, which are listed in Table~\ref{tab2}. For these cross
section calculations of $gg\to(n-2)\,g$ scattering processes at a
center-of-mass energy of 14 TeV, we have used the CTEQ6L1 parton
density function set~\cite{Pumplin:2002vw} as implemented in
LHAPDF~\cite{Whalley:2005nh} with a fixed renormalization and
factorization scale taken at $M_Z=91.188$ GeV. For the jet cuts, we
have chosen $p_T^\mathrm{jet}>20$ GeV, $|\eta^\mathrm{jet}|<2.5$ and
$\Delta R_\textrm{jet--jet}>0.4$. The cut efficiencies for different
numbers $n$\/ of gluons can be read off Table~\ref{tab2}. Employing
these cuts we were also able to verify the jet
production cross sections that we have produced using {\sc Comix} with
the results reported in Ref.~\cite{Gleisberg:2008fv}. To have a stringent
comparison, we ran {\sc Comix} for pure gluon scatterings yielding
cross sections that take the full-color dependence into account. These
results are also listed in the table; for the 4-gluon and 5-gluon
processes, they can be directly compared to the cross section
estimates obtained with the {\sc Tess} MC on the GPU, since the
leading-color approximation already gives the exact result. The
agreement is found to be satisfactory.

We show differential distributions in Figure~\ref{fig3}. To obtain
them we again used $10^9$ sweeps where, for a certain gluon
multiplicity, the total number of generated events can be read off
Table~\ref{tab2}. We kept most of the input parameters unaltered
except for the jet cuts, which we changed to $p_T^\mathrm{jet}>60$ GeV,
$|\eta^\mathrm{jet}|<2.0$ and $\Delta R_\textrm{jet--jet}>0.4$,
and the choice of the renormalization and factorization scales, which
we decided to set dynamically using $H_T$ as a scale. On the right
hand side of Figure~\ref{fig3} we show for 3, 5, 7, 9 gluon jets in
the final state the normalized distributions for the $H_T$ observable
and the minimum $R$-separation, $R_\mathrm{min}$, which we define
through the jet--jet pair being closest in $R$-space,
$R_\mathrm{min}=\min\{\Delta R_{ij}\}$. As can be seen smooth
distributions are easily obtained using the {\sc Rambo} phase-space
generator. They are normalized to the total cross sections, which have
been calculated by {\sc Tess} as
$\sigma_5=(6.97838\pm0.00044)\times10^4$~pb,
$\sigma_7=(4.9761\pm0.0043)\times10^2$~pb,
$\sigma_9=(4.532\pm0.044)$~pb and
$\sigma_{11}=(4.51\pm0.19)\times10^{-2}$~pb.
On the left hand side of Figure~\ref{fig3} we have added profile plots
displaying the relative gauge invariance versus the decimal logarithm
of the matrix-element weight. Specifically, we show the average 
$\left|K[1]\cdot\,J[2,\ldots,n]\right|^2 /
\left|J[1]\cdot\,J[2,\ldots,n]\right|^2$ and its mean standard
deviation as a function of the matrix-element weight
$W_\mathrm{ME}=\left|m(1,2,\ldots,n)\right|^2=
\left|(J[1]\cdot\,J[2,\ldots,n])\times K^2[2,\ldots,n]\right|^2$.
The behavior is as expected; for large weights, we see gauge
cancellations up to float precision. For small weights, the gauge
cancellations are less precise. However, these small-weight events are
not important since they do not contribute to the calculation of the
observables.

\FIGURE[!t]{
\label{fig4}
\centerline{
  \includegraphics[clip,width=1.0\columnwidth]{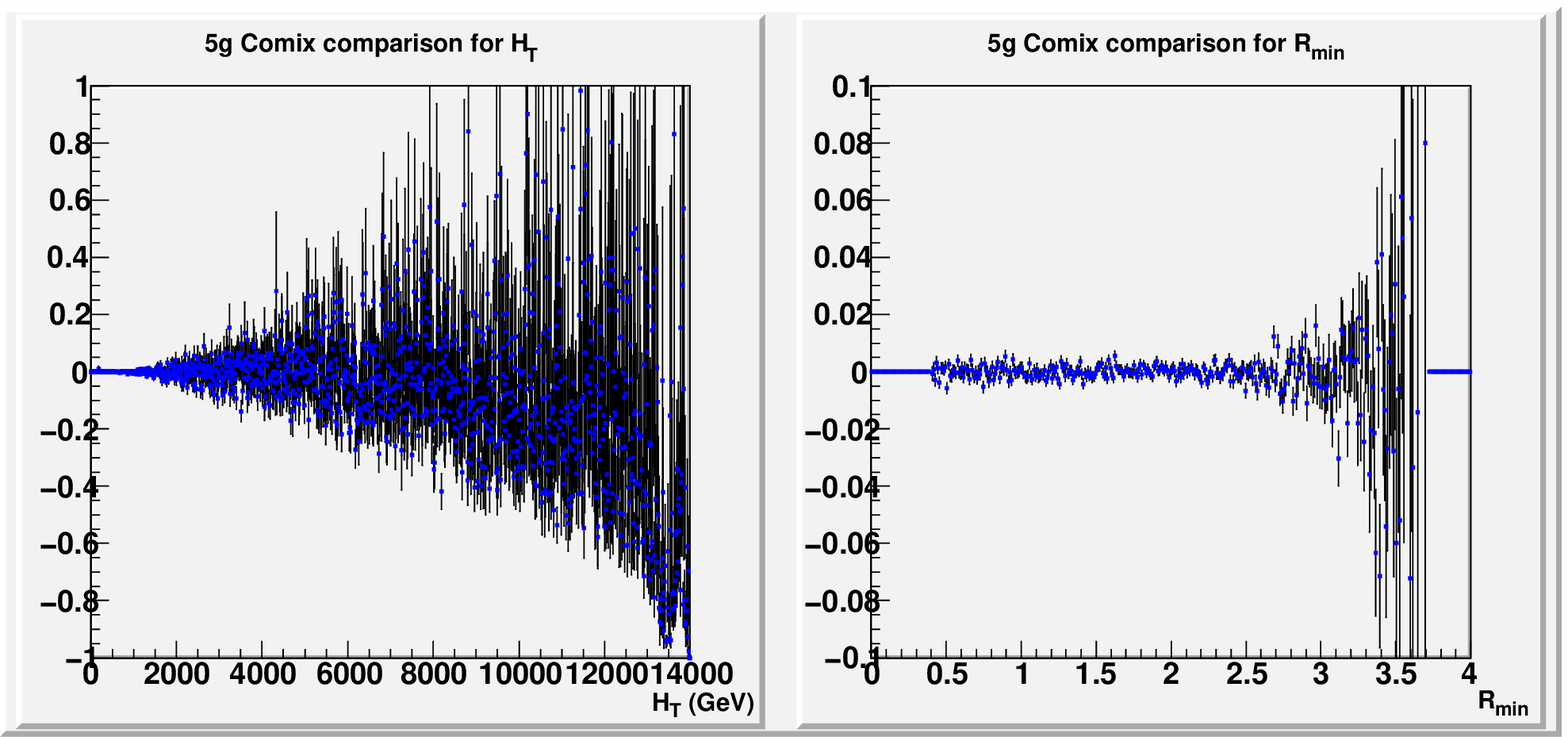}}
\caption{
The ratio $(\sigma_5^\textsc{Tess}\times d\sigma_5^\textsc{Comix}/dX) /
(\sigma_5^\textsc{Comix}\times d\sigma_5^\textsc{Tess}/dX)-1$
for the 5-gluon $X=H_T$ (left panel) and $X=R_\mathrm{min}$
distributions, see the text.
The mean standard deviation error bars of the {\sc Comix} calculation
are also shown.}}

Finally, in Figure~\ref{fig4} we compare our results to the results
obtained with the {\sc Comix} event generator \cite{Gleisberg:2008fv}.
For this comparison, we use both the $H_T$ and $R_\mathrm{min}$
5-gluon distributions and we fix the renormalization and factorization
scale through $M_Z=91.118$ GeV to avoid any issues resulting from slight
differences in the evolution codes for running scales between the two
MCs. Furthermore, to have a sole shape comparison, we plot the ratio
$(\sigma_5^\textsc{Tess}\times d\sigma_5^\textsc{Comix}/dX) /
(\sigma_5^\textsc{Comix}\times d\sigma_5^\textsc{Tess}/dX)-1$ with the
results shown in Figure~\ref{fig4} and $X$ being the observable in
consideration. Note that for the minimum $R$-separation distribution,
we have excellent agreement with {\sc Comix}. For the $H_T$
distribution, we have to realize that the cross section spans 28
orders of magnitude. As {\sc Comix} relies on importance sampling, it
only sparsely populates the tail of the distribution. This leads to
large uncertainties at large values of $H_T$ and, in these regions,
{\sc Comix} will hence tend to underestimate the value for the cross
section.

\section{Conclusions and Outlook}
\label{sec:concl}

In our first exploration of the potential of using multi-threaded GPU-based
workstations for Monte Carlo programs, we obtained very encouraging results.
We implemented the entire {\sc Tess} Monte Carlo on the GPU chip; the
only off-chip usage occurs through utilizing the texture memory for
the evaluation of the parton density function and the strong coupling
constant. The GPU global memory is solely used for transferring the
Monte Carlo results to the CPU memory. At this exploratory phase of
the project, we limited ourselves to the calculation of leading-color
leading-order $n$-gluon matrix elements. With respect to the CPU-based
implementation of our Monte Carlo we have found impressive
speed-ups in the computations reaching from ${\cal O}(300)$ for
$PP\to2$ jets to ${\cal O}(150)$ for $PP\to10$ jets.

Given these results we are encouraged to further develop the {\sc Tess}
Monte Carlo by including quarks, vector bosons and subleading color
contributions. We are also planning to implement on the GPU a
dipole-based phase-space generator like {\sc Sarge} as an alternative to
the unit-weight phase-space generator {\sc Rambo}. This will avoid
the under-sampling issues in high jet-multiplicity final states. 
These improvements will result in a full leading-order parton-level
event generator, which will be at least two orders of magnitude faster
than existing leading-order parton-level generators.

More importantly, a GPU-based Monte Carlo can be used as the generator 
for the real corrections in an automated next-to-leading order
parton-level MC generator. The virtual corrections can be calculated by using 
a generalized-unitarity based method. These methods themselves spend about
90\% of their evaluation time calculating leading-order matrix elements.
By using the GPU-based matrix-element evaluator of the {\sc Tess} Monte Carlo,
one can safely estimate a speed-up factor of ${\cal O}(10)$ in the
evaluation time of the virtual corrections. This means that the
GPU-based automated NLO parton-level generator can be successfully
implemented on the GPU-based workstation and obtain accurate results
on a timescale of a day without resorting to a large-scale PC farm.

Finally, GPU chips for numerical evaluations are still evolving
rapidly. This will lead to additional significant speed-ups over
CPU-based Monte Carlos in the coming years.

\acknowledgments

We want to thank Jim Simone for suggesting the Tesla GPU chips for use
in event generators. We thank the High-Performance Computing Department
at Fermilab for giving us support and access to the LQCD Tesla-based
workstations.

We would also like to thank Patrick Fox, who came up
with the name ``{\sc Tess}'' and suggested it to us as the name for
our Monte Carlo program.

Fermilab is operated by Fermi Research Alliance, LLC, under contract
DE-AC02-07CH11359 with the United States Department of Energy.

%-----------------------------------------------------------------
%\newpage

\newpage

\end{document}